# Methodology for Holistic Reference Modeling in Systems Engineering

Dominik Ascher, Erik Heiland, Diana Schnell, Peter Hillmann, and Andreas Karcher

*Universität der Bundeswehr München, Werner-Heisenberg-Weg 39, 85577 Neubiberg, Germany*

**Abstract**
Models in face of increasing complexity support development of new systems and enterprises. For an efficient procedure, reference models are adapted in order to reach a solution with les overhead which covers all necessary aspects. Here, a key challenge is applying a consistent methodology for the descriptions of such reference designs. This paper presents a holistic approach to describe reference models across different views and levels. Modeling stretches from the requirements and capabilities over their subdivision to services and components up to the realization in processes and data structures. Benefits include an end-to-end traceability of the capability coverage with performance parameters considered already at the starting point of the reference design. This enables focused development while considering design constraints and potential bottlenecks. We demonstrate the approach on the example of the development of a smart robot. Here, our methodology highly supports transferability of designs for the development of further systems.

**Keywords**
Reference Modeling, Enterprise Architecture, Methodology

## 1. Introduction

Enterprise Architecture (EA) provides holistic and model-based methods to describe capabilities, operational scenarios, services, systems and technology and their underlying functionality necessary to perform given operational missions. According to IEEE 1471 [1], an architecture describes the *"fundamental organization of a system, embodied in its components, their relationships to each other and the environment, and the principles governing its design and evolution"*. EA is related to Model-based Systems Engineering (MBSE), i.e. the formalized application of models for system development activities [2], helping tractability of system complexity, quality improvement as well as reuse of information. Major differences between EA and MBSE are intended scope and model structure, which are expressed by architecture frameworks. According to the ISO 42010 Standard [3], an architecture framework embodies the "conventions, principles and practices for the description of architectures established within a specific domain of application and/or community of stakeholders". Reichwein [4] consider architecture frameworks to comprise a definition of structure and content of architecture descriptions and incorporate best practices to establish the latter. Then, an architecture framework can be considered a basic foundation on which an architecture description can be established.

The Open Group defines with their architecture framework (TOGAF) itself as a *"foundational structure, or set of structures, which can be used for developing a broad range of different architectures"* [5]. In addition, they deem it to comprise method as well as a set of building blocks, a set of tools, common vocabulary and a list of recommended standards and compliant products to implement the building blocks. MacKenzie et al. [6] define a reference model as *"an abstract framework for understanding significant relationships among the entities of an environment, and for the development of consistent standards or specifications supporting that environment"*. Consequently, reference models and reference architectures are abstract solutions for the design of systems in a specific domain, where these solutions are also known as patterns. More specifically, a reference architecture consists of reusable models and patterns, which are customizable. It describes a particular recurring design problem for specific design contexts and presents a well-proven solution for the problem [7].

In this paper, we present a holistic approach for deriving and utilizing a reference architecture. Here, it combines operational-level knowledge about intended operational scenarios, actors and the environment the latter operate in, to analyze and design domain-specific model configurations. Benefits include an end-to-end traceability of the capability coverage as well as flexible transferability of designs for the development of further systems between different domains by providing reusable reference model building blocks.

This paper is structured as follows: In Section 2, we describe an example problem, derive requirements for a comprehensive approach and describe our research questions. Section 3 describes the current state of the art for our problem. Our contribution in Section 4 elaborates on our approach for reference modeling. Then, we demonstrate and evaluate our approach for the initially defined problem in Section 5. Finally, we conclude and provide an outlook on future work.

## 2. Scenario and Requirements

We consider our problem scenario to be as follows: *MyGarden* is a medium-sized company that specializes in covering the service needs of the garden sector in the Business to Customer (B2C) segment. MyGarden provides a wide range of garden-specific services to customers such as mowing, scattering and weed removal services. Therefore, MyGarden employs staff and holds inventory of its own gardening tools. Recently, company management has decided to focus on promising new business models and technologies such as *"Smart Gardening"*, where classic employee-intensive gardening services are supported by autonomous systems such as specialized robots. These new systems are developed by the R&D department on a modular basis using the systematic approach of reference models. A major challenge is the company's requirements management. Changed requirements imply that only limited use can be made of existing service descriptions, while these are necessary due to dependencies with internal and external partners. The company's managers also wish for transparency of capabilities specified for MyGarden's robots and offered services for communication to externals.

To address these challenges, a company-wide EA should be established, which provides an architectural modeling framework for mapping corresponding organization business and IT structure. As a basis for the application of MBSE, modeling using the meta-model of the *Unified Architecture Framework* (UAF) [8] is performed. In a second step, reference models for according robots are established for the Smart Gardening business area, which will serve to promote usability and adaptability for three robot variants. To this end, core functionalities such as automatic obstacle detection and objects recognition, localization, and the execution of systematic robot action strategies are to be abstracted on the basis of services and mapped to corresponding reference model modules. These are flexibly exchangeable as well as combinable through standardized interfaces. Reference modeling can simplify capability management, optimize product lines, reduce costs and increase service quality by merging business, service and product-relevant characteristics. As a result, the following requirements for an approach for reference modeling have been identified:

- **R1**: The approach shall support development of architecture descriptions by providing reusable assets helping adaptation and parameterization of the reference model to individual domains.
- **R2**: The approach shall provide guidance for using reference architecture descriptions as well as communicating them to relevant stakeholders.
- **R3**: The reference architecture shall provide a reusable architecture description for different parts of modular systems.

- **R4**: The reference architecture shall offer traceability of model elements from intended capabilities, to operational scenarios, over services to individual systems and help visualize dependencies between them.
- **R5**: The reference architecture should support standardization and make use of existing, prevalent architecture framework standards such as the Unified Architecture Framework (UAF).
- **R6**: The reference architecture should support architecture decisions in terms of selecting the most suitable target architecture configuration for a given domain from a set of possible alternatives.

Based on the aforementioned requirements, we consider the following research questions to be important:
- **RQ1**: How can a methodology support application of a reference architecture parts to different domains by encouraging reusability and transferability?
- **RQ2**: How can reference architecture be used for model-based description from strategic capabilities, over operational scenarios, requirements and provided services to technical implementation within individual systems as well as provide traceability between these concepts throughout different levels of model hierarchy?
- **RQ3**: How can building blocks and patterns of a reference model be adopted, adapted and extended for individual application domains?

## 3. Related Work

Reference models define common methodology and elements, which can be readily adapted for different application contexts or domains, instead of defining these anew, when constructing domain-specific application models. Reference models are specifically concerned with the issue of reuse of models, which is recognized in international EA [9] and MBSE literature [10], for instance in the context of configuration management [11] or, more generally, software [12] and knowledge patterns [13]. Thus, we deem reference modeling to be concerned with models reusable for the construction of individual model configurations. The application of a reference model to a particular domain-specific or application model is achieved using different mechanisms. These include (1) adoption, i.e. the direct usage of reference model elements, (2) adaptation, i.e. the modification and tailoring of reference model elements and using them in the application model, as well as (3) extension, i.e. the extension of existing reference model elements by supplementing domain-specific information for constructing an application model [14].

In terms of existing EA frameworks, several approaches define reference models as a central concept. For instance, the *Federal Enterprise Architecture Framework* (FEAF) [15] is used for describing U.S. federal government agency management of IT-landscapes using as hared enterprise architecture. For this, a key element is the consolidated reference model (CRM), which defines five reference models helping IT-landscape management concerning business, service, components, technical as well as data. However, the FEAF is mainly suited for governmental agencies and not related to our domain. Instead, TOGAF describes reference libraries containing reference architectures and reference models, where *"a generic reference architecture provides the architecture team with an outline of their organization-specific reference architecture that will be customized for a specific organization"* [5]. Here, reference architectures can be differentiated in terms of foundation architectures, common systems architectures as well as industry and organization-specific architectures. In addition, TOGAF defines the *Technical Reference Model* (TRM) as well as the *Integrated Information Infrastructure Reference Model* (III-RM). Finally, *Service Oriented Architecture* (SOA) represents a *"paradigm for organizing and utilizing distributed capabilities that may be under the control of different ownership domains"* [6]. Oliveira et al. [16] provide a literature review over numerous reference models and reference architectures based on SOA. For instance, the OASIS Reference Model for SOA [6] provides a template for the elements and relationships necessary for realizing a SOA, while abstracting from concrete technologies. The Open Group published a specification for a SOA Reference Architecture (SOA RA) [17], which can be instantiated using TOGAF [18]. While defining structure for reference models, we found that the considered EA frameworks lack concrete details about how reference models are implemented.

Finally, we consider (reference) models and architectures for robotic systems. Hayes-Roth et al. [19] and Albus et al. [20] propose reference architectures for intelligent systems design. The latter consider a reference model architecture, which describes a hierarchy comprised of layers for individual processing modules such as behavior generation, world modeling as well as sensory processing, value judgment and knowledge databases, which are connected using communication pathways between them.

However, while providing an extensive architecture description, they mainly focus on system-level description. Feitosa et al. [21] and Ahmad et al. [22] provide overviews of (reference) architectures for robotic systems. In this context, the latter provide a mapping study for software architectures in particular, and found that service-oriented architectures supporting an interconnected web of robots to represent emerging solutions. While providing service-oriented system details, we found that none of the existing reference models and architectures for robotic systems to provide a comprehensive reference architecture from strategic, over operational and service, to system modeling levels in an EA context. Therefore, we found that none of the considered approaches fully suit our requirements and comprehensively address our research questions.

## 4. Concept for Reference Modeling

In this section we describe our holistic approach employed for reference modeling.

### 4.1. Overview

Reference models provide reusable model building blocks and patterns for single or a set of particular (application) domains. In the following, we describe various modules of our reference modeling methodology supporting aforementioned purpose. For this, Figure 1 provides an overview of our methodology. It consists of different viewpoints from the highest strategic levels to the detailed descriptions of the traceability, connecting the different levels. The reference modeling methodology is comprised of the following elements:

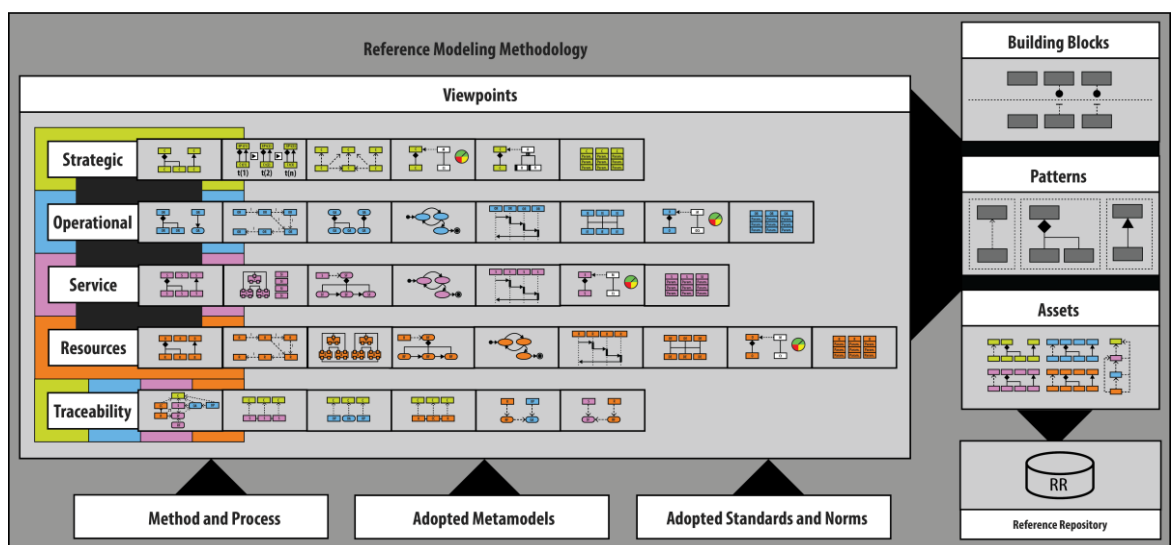

**Figure 1**: Overview of reference modeling methodology.

**Viewpoints** are defined by their subjects of concerns such as Strategic, Operational, Service or Resources as well as their aspects of concerns, such as Structure, Behaviors, Parameters or Requirements. Here, our reference model is composed of layers describing different aspects and subjects of concern, which are defined using specific viewpoints. In the ISO 42020 Standard [23], a viewpoint represents the *"architecture conventions for the construction, interpretation and use of architecture views to address specific concerns about the architecture entity"*, where a concern relates to a *"matter of interest or importance to a stakeholder"*. Figure 2 shows the different concerns and their modeling notation for different viewpoints. Note that we utilize of a subset of UAF viewpoints [8], we deem most relevant for our reference modeling approach and its application, but additional or customized viewpoints could be utilized for application as well, depending on considered meta-models, standards as well as processes.

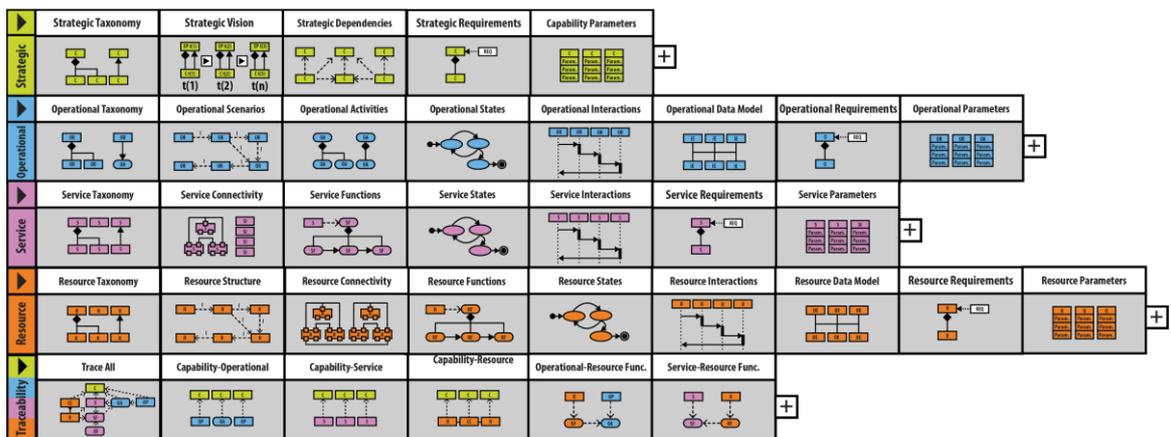

**Figure 2**: Relationship between concerns, modeling notations and viewpoints.

**Views** describe individual perspectives on the model itself based on viewpoints and are constructed using assets. According to the ISO 42020 Standard [23][23], an architecture view represents an *"information item expressing the architecture from the perspective of specific stakeholders regarding specific aspects of the architecture entity and its environment"*, where the architecture entity refers to the *"thing being considered, described, discussed or otherwise addressed during the architecting effort"*, i.e., in our context, the reference model. From the latter, assets and viewpoints are utilized to construct views using a process and based on a meta-model and notation.

**Assets** refer to the total set of all model-related elements belonging to the reference model such as building blocks, patterns as well as individually created architecture viewpoints and views. Combined model elements, which make up the reference model, are stored in the reference repository.

**Building Blocks** refer to the model elements themselves and their connections between them. In this, single building blocks can be connected to each other using typed interfaces. Thus, interfaces can then be defined as providing or as requiring a building block of a specific type as well as between different concern levels. The Open Group defines building blocks as potentially reusable components of business, IT or architectural capability, which can be combined with other building blocks to deliver architectures or solutions [5]. Figure 3 provides an overview of building blocks, their interfaces and exemplary application. In terms of notation, bowls denote required interfaces, whereas circles refer to provided interfaces of a given type for building blocks. Part A shows building blocks for different concerns, whereas Part B shows an exemplary building block configuration, which visualizes traceability throughout different concern levels. As building blocks can be switched, Part C shows different configuration alternatives and, finally, Part D shows the internal structure of building blocks for exemplary resources.

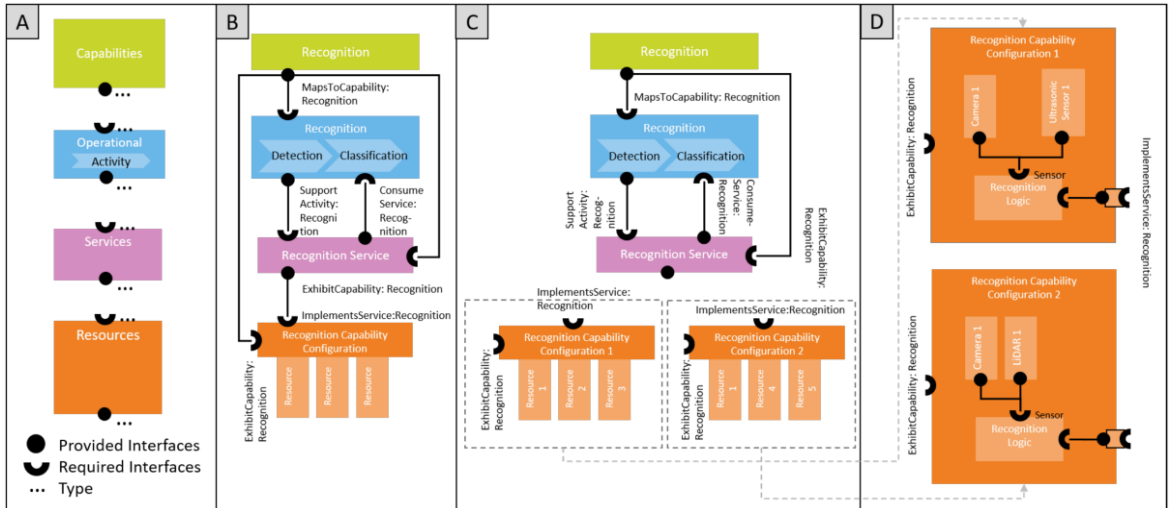

**Figure 3**: Representation of building blocks.

**Patterns** are described as formations between different building blocks. As predefined sets of building blocks, they are intended to describe reusable configurations of different building blocks. Bass [24] define an architectural pattern to be a description of element and relation types with a set of constraints on how they may be used. Then, according to them, a pattern can be considered a set of constraints on an architecture as well as on the element types and their patterns of interaction and constraints define a set or family of architectures that satisfy them. The Open Group defines a pattern to be the following [5]: *"patterns are considered to be a way of putting building blocks into context; for example, to describe a re-usable solution to a problem"*. Figure 4 shows an exemplary application of patterns, representing individually composed sets of building blocks. Here, a service pattern is applied to an existing model consisting of strategic and resource model elements. Pattern application results in the merge of the existing model elements with the new services.

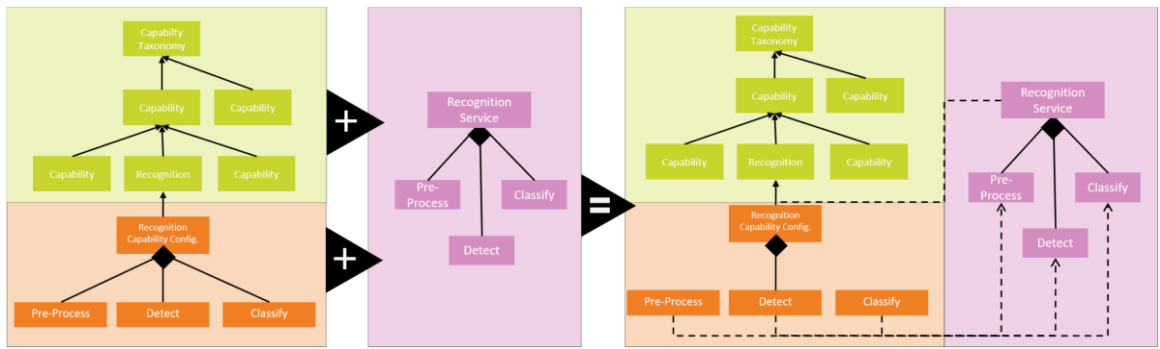

**Figure 4**: Representation of exemplary pattern application.

**Reference Repository** describes a collection of all model-related artifacts belonging to the reference model, which defines structure and format as well as location how this collection is stored [25]. In the ISO 42020 Standard [23], a repository refers to a *"place where work products and the associated information items are or can be stored for preservation and retrieval"*. In our context, these work products and information items relate to the reference model itself and related artifacts resulting from its process and methodology. The Open Group considers architecture repositories to comprise the architecture landscape, reference libraries, architecture capabilities, standards, governance as well as architecture method and meta-models [5].

Method and Process describe how the reference model should be applied and for this cause provide a set of action strategies for constructing a domain-specific model configuration based on the reference model. Here, ISO 42020 Standard [23] defines a process as a *"set of interrelated or interacting activities that transforms inputs into outputs"*. For instance, in our context, inputs could be represented by the reference model and its model-related artifacts on the one hand side and domain-specific modeling requirements on the other hand side. Instead, the outputs would be a set of domain-specific model configurations created with the reference model, which address aforementioned requirements, and therefore are created by transforming the inputs.

## 4.2. Levels of Concerns

The central element of the **strategic** concerns is the capability. Capabilities describe abilities or competencies, which are achieved from a combination of business or operational processes, performers, services and resources. We describe strategic concerns in terms of structure such as structural taxonomies, dependencies and future planning (i.e. strategic visions) as well as requirements and parameters. Further note that capability concerns do not describe behavioral aspects.

**Operational** concerns describe operational processes of an organization as well as the actors that perform them. In terms of their description, they can be considered business and mission scenarios or business processes. More specifically, *operational activities* specify the individual steps with in operational processes, where each step can be assigned to an *operational performer*, i.e. an entity that executes them. We describe operational concerns in terms of structure such as structural taxonomies dependencies, behavioral functionality such as performance of operational activities, behavioral states as well as interactions, conceptual data models, requirements as well as parameters.

**Service** concerns describe the technology independent functionality necessary for performing specific tasks. Here, services describe functionality independent of implementation. TOGAF describes a service as *"an element of behavior that provides specific functionality in response to requests from actors or other services"* [5]. We describe service concerns in terms of structure such as structural taxonomies, dependencies and connectivity, behavioral functionality, behavioral states as well as interactions, requirements and parameters.

**Resource** concerns describe the resources such as systems and system configurations, which realize functionality in terms of system-level implementation. Therefore, they describe how resources are configured and connected to exhibit capabilities and implement service functionality. We describe resource concerns in terms of structure such as structural taxonomies, dependencies and connectivity, behavioral functionality, behavioral states as well as interactions, physical data models, requirements and parameters.

Finally, **traceability** concerns describe the relationships between different concerns. Our approach is capability-centered, i.e. capabilities are realized by individual concerns such as operational activities, operational performers, services as well as resources and resource configurations. Therefore, we describe these realization and exhibition relationships for the relevant concepts and their corresponding capabilities.

## 5. Assessment

Subsequently, firstly we demonstrate application of reference modeling on a small example scenario for smart robots as defined in Section 2. Secondly, we evaluate how holistic models can support possible simulation as well as optimization for system-level decisions.

## 5.1. Demonstration

As previously established, reference models provide reusable model building blocks and patterns for single or a set of particular (application) domains. Figure 5 provides an overview of this domain application.

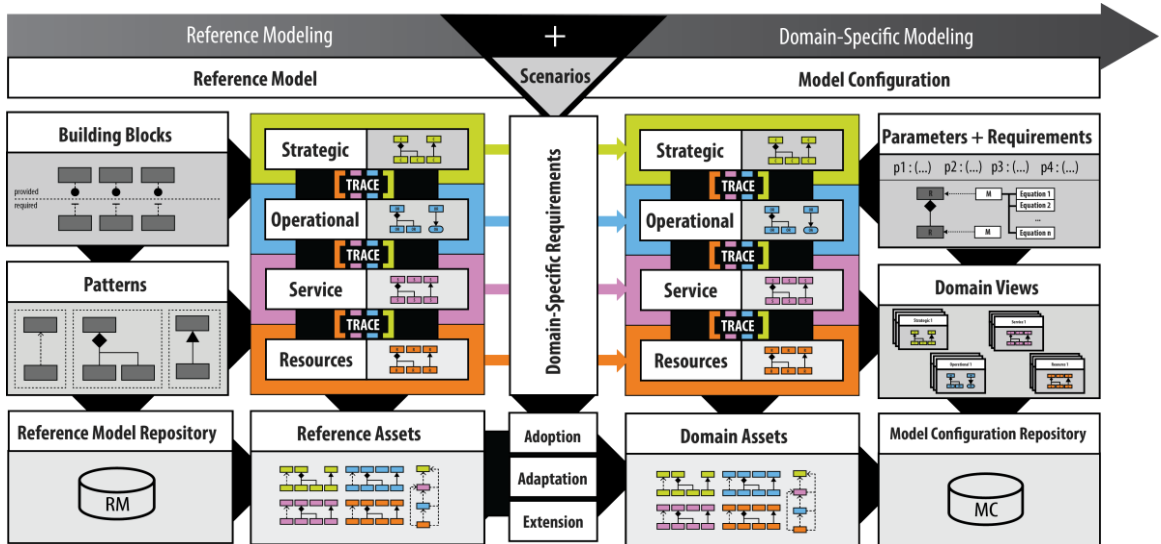

**Figure 5**: Representation of reference model and application to domain.

For demonstration of the application of our reference modeling methodology, we utilize the meta-model of the UAF (UAF-DMM) for modeling. Note that other modeling languages and their defined meta-models such as ArchiMate [26] could be used for modeling using our methodology as well. Based on defined concerns, we describe the individual domain-specific model elements for our example scenario, which is modeled using the reference model. Due to the limited extent of this paper, we describe the concerns in terms of a set of domain-specific model elements only. This set could then be further broken down into individual views (see Figure 3) defining these concerns, which would then represent individual subsets of the total set of described model elements. Figure 6 shows the resulting model elements as well as the reference model elements they are derived from in the upper right corner of individual model elements, which can be distinguished as follows:

- **Strategic**: The model defines capabilities for the mowing robot, which contains capabilities for recognition, mobility as well as mowing.
- **Operational**: The model defines an operational performer (mowing node), which performs various activities such as recognize mowing object, moving in green areas as well as the mowing process.
- **Service**: For services, a smart mowing service is defined, which contains individual services for object recognition, green area mobility, as well as mowing itself.
- **Resource**: In terms of resources, the model defines a capability configuration for the mowing robot, which contains sensors such as a camera as well as actors such as batteries, mowing blades and the propulsion system. The configuration is capable to perform various functions such as pre-processing, detecting and classifying.
- **Traceability**: Between the different model elements, relationships between the different modeling layers are defined using traceability relations prescribing exhibition and mapping to capabilities, as well as performance and implementation of functions and activities.

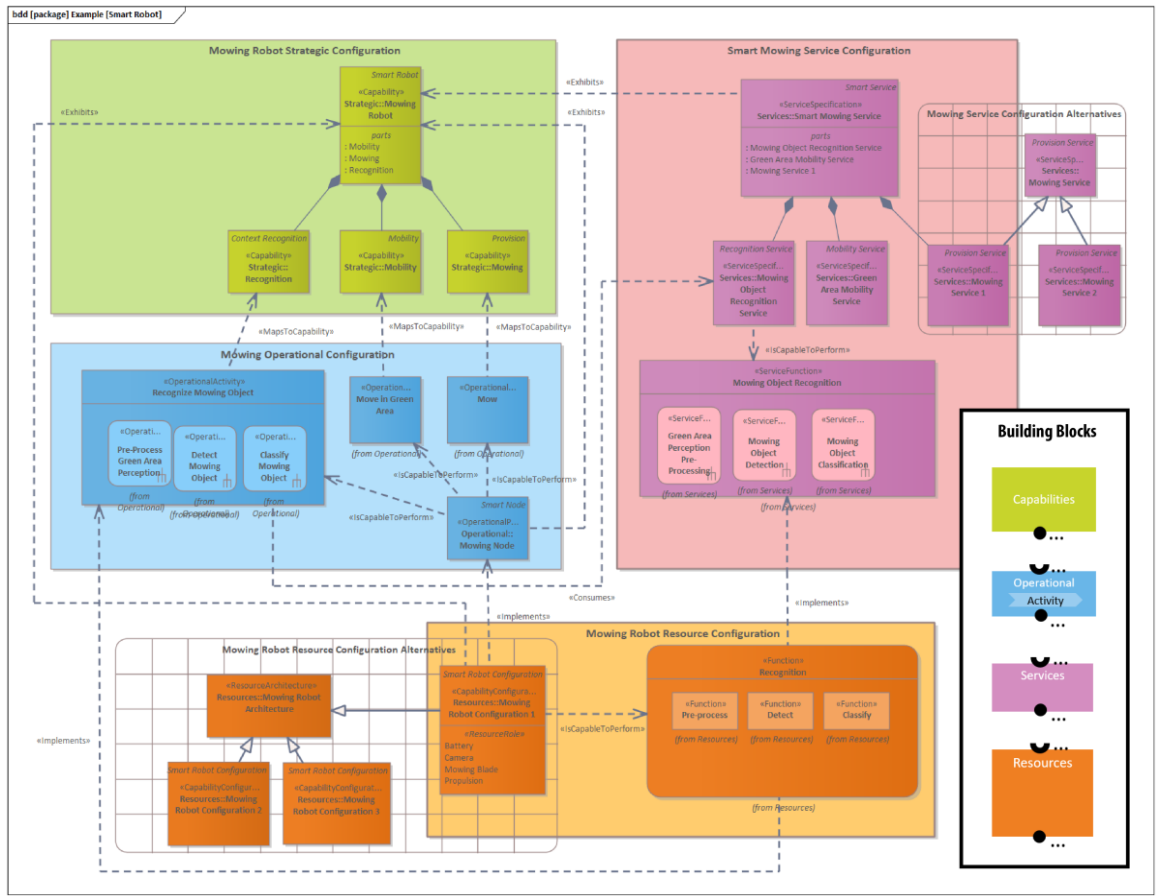

**Figure 6**: Application of the reference model to the domain-specific smart robot example.

## 5.2. Evaluation

Based on the demonstration, we describe evaluation of our reference modeling approach in terms of a concrete evaluation for our example scenario defined in Section 2. We focus on describing energy consumption for different system configurations as an example in terms of how the high-level requirement of the longest possible battery life affects the nature of individual system components and behavior. The range of a battery-powered vehicle depends on many factors. For instance, these include its weight and battery capacity, as well as environmental conditions such as characteristics of the traversed terrain. For the following simulation, we assume a simplistic system model, where the weight and capacity of the device are fixed values. We try to optimize our system to drive as efficiently as possible over a given terrain profile. To do this, we simulate the system's power consumption based on two different algorithms. Here, algorithms are realized as exchangeable building blocks in the reference model.

Figure 7 shows a simplified model to illustrate the impact of terrain on electricity consumption. A matrix can be created for each terrain by dividing it into grids and assigning a number to each area, starting from an initial elevation, indicating whether the area is the same elevation, lower, or higher. For this, in our example, we assign values from 0-3. Depending on whether the next field to be approached is higher, lower or equal elevation, the power consumption of the device will change.

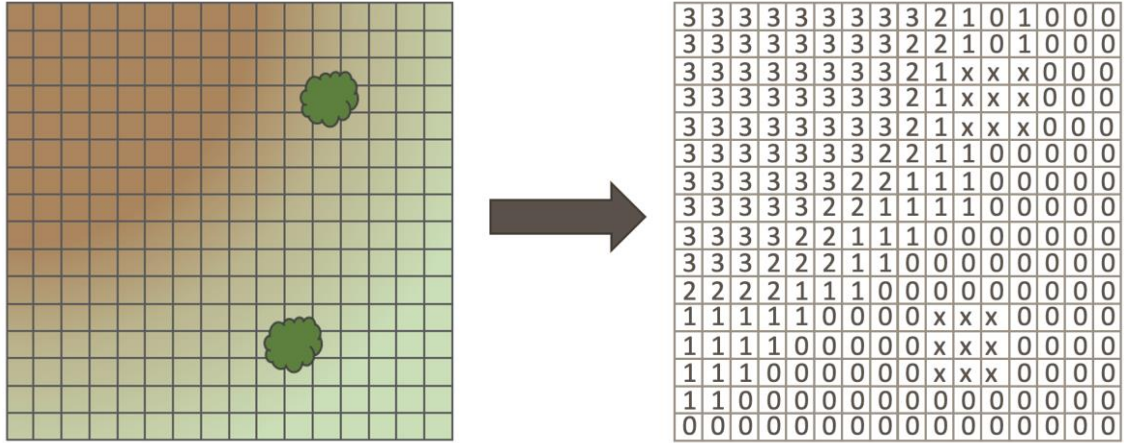

**Figure 7**: Generation of a matrix for the elevation profile of a terrain.

Therefore, we assume the following values for each position change p:

$$p_{t+1} - p_t \begin{cases} > 0 \rightarrow 1.9 \text{ (high)} \\ = 0 \rightarrow 1.0 \text{ (normal)} \\ < 0 \rightarrow 0.6 \text{ (low)} \end{cases} \qquad (1)$$

The mowing algorithm is a system function of the lawn mower control unit. Two algorithms have been developed to determine the path: Algorithm 1 finds its path by subsequently traversing the map and orienting itself only to edge points, obstacles and the areas already passed. Instead, Algorithm 2 additionally considers the terrain profile and tries to keep the elevation difference to the next field as small as possible at high points. Based on elevation for each field, we then determine according energy consumption as h (high), n (normal) and l (low). Here, simulation of the power consumption is then performed within the battery component. This is declared as a *SysML* block and contains a parametric diagram in which the system behavior is modeled (see Figure 8).

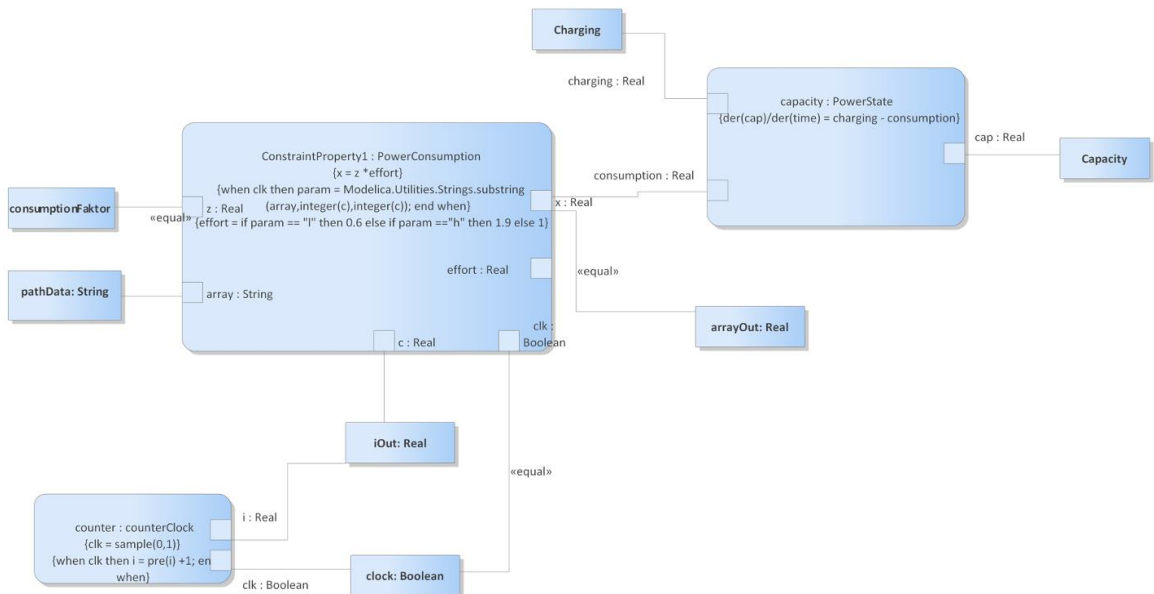

**Figure 8**: Parametric Diagram for the calculation of the power consumption.

Three constraints were included in the model. The *PowerState* constraint calculates the remaining amount of current, depending on the consumption and the supplied charging energy. Power consumption is calculated by the *PowerConsumption* constraint based on path data. Then, timing is controlled by the *counterClockcomponent*. The consumption factor is used with in this model to scale the power consumption. In a complete model, this could include additional factors, such as vehicle weight and speed.

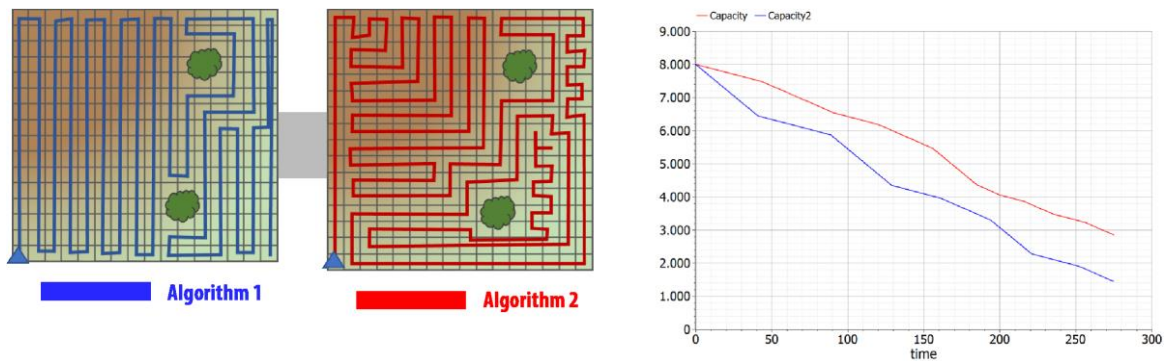

**Figure 9**: Resulting paths (left) and power consumption (right) for two different algorithms.

The map shown in Figure 7 is then solved by the two algorithms we initially discussed. Figure 9 shows the result of this simulation. For the given map, the second algorithm is more efficient. However, this is not necessarily the case for all terrain profiles. Simulation with different maps could for instance lead to the result that in most cases, the efficiency of one algorithm prevails and should be used for the system. Alternatively, more adaptive selection could also be applied, which chooses one of the algorithms depending on the given context. The experiment partly demonstrates the influence of highly aggregated requirement profiles on the design of system components. To explore systems holistically, more comprehensive consideration of the problem could involve a larger number of system parameters that influence each other and could be differently parameterized, while resulting in higher problem complexity and dimensionality.

In summary, all blocks have clear defined connections, so we can easily exchange parts of the entire system. This becomes only possible through a stringent usage of reference modeling with well-defined interfaces and data formats for exchange. In this way, the system design becomes explorable like plug-and-play. So, we can easily evaluate different system designs, configurations and setups.

## 6. Summary and Outlook

In this paper, we presented a holistic approach for reference modeling to describe reference models across different views and levels. We presented a methodology to support traceable reference modeling to analyze and design domain-specific model configurations. Here, proposed benefits include an end-to-end traceability of the capability coverage as well as transferability of designs for the development of systems between different domains. For this, we demonstrated modeling a small example problem for smart robots and evaluated how holistic models can support possible simulation as well as optimization for system-level decisions. As the presented example of smart robots represents a specific domain, for future work, it has to be investigated how the modeling approach can be applied to additional domains as well as how domain knowledge can be further abstracted within the reference model, while conforming with domain-specific standards and applicable meta-models.

# 7. References

[1] IEEE, *Recommended Practice for Architectural Description for Software-Intensive Systems*, IEEE1471, 2000.
[2] INCOSE SE, *Vision 2020*, Technical Report (INCOSE-TP-2004-004-02), 2007.
[3] ISO42010, *Systems and software engineering: Architecture description*, ISO/IEC/IEEE 42010, 2011.
[4] A. Reichwein, C. Paredis, *Overview of architecture frameworks and modeling languages for model-based systems engineering*, ASME 1275, 2011.
[5] The Open Group, *TOGAF Version 9.1*: Open Group Standard, TOGAF, 2011.
[6] C.M. MacKenzie, K. Laskey, F. McCabe, P.F. Brown, R. Metz, B.A. Hamilton, *Reference model for service-oriented architecture 1.0*, OASIS Standard, 2006.
[7] F. Buschmann, K. Henney, D.C. Schmidt, *Pattern-oriented software architecture: on patterns and pattern languages*, vol. 5 of Wiley Series in Software Design Patterns, John Wiley and Sons, 2007.
[8] Object Management Group, *Unified Architecture Framework (UAF)*, 2020.
[9] P. Bernus, *Enterprise models for enterprise architecture and ISO 9000*, Annual Reviews in Control 27(2), 211-220, 2003.
[10] E.R. Carroll, R.J. Malins, *Systematic Literature Review: How is Model-Based Systems Engineering Justified?*, 2016.
[11] A.M. Madni, M. Sievers, *Model-based systems engineering: Motivation, current status, and research opportunities,* Systems Engineering 21, 172–190, 2018.
[12] D.C. Schmidt, M. Fayad, R.E. Johnson, *Software Patterns*, Communications of the ACM 39(10), 37–39, 1996.
[13] A. Sutcliffe, *The Domain Theory: Patterns for knowledge and software reuse*, CRC Press, 2002.
[14] J. vom Brocke, C. Buddendick, *Konstruktionstechniken für die Referenzmodellierung: Systematisierung, Sprachgestaltung und Werkzeugunterstützung*, in J. Becker, P. Delfmann, (Eds.), Referenzmodellierung, Physica, Heidelberg, pp. 19–49, 2004.
[15] Office of Management and Budget, *Federal Enterprise Architecture Framework (FEAF)*, 2012.
[16] L.B.R. de Oliveira, K.R. Felizardo, D. Feitosa, E.Y. Nakagawa, *Reference Models and Reference Architectures based on Service-oriented Architecture: A Systematic Review*, in: European Conference on Software Architecture, Springer, pp. 360–367, 2010.
[17] The Open Group, *Service Oriented Reference Architecture (SOA RA)*, TOGAF, 2009.
[18] The Open Group, Using TOGAF to Define and Govern Service-Oriented Architectures, TOGAF Series Guide, 2011.
[19] B. Hayes-Roth, K. Pfleger, P. Lalanda, P. Morignot, M. Balabanovic, *A domain-specific software architecture for adaptive intelligent systems*, IEEE Transactions on Software Engineering 21, 288–301, 1995.
[20] J.S. Albus, *RCS: A reference model architecture for intelligent systems*, in: Working Notes of AAAI Spring Symposium on Lessons Learned for Implemented Software Architectures for Physical Agents, pp. 1–6, 1995.
[21] D. Feitosa, E.Y. Nakagawa, *An Investigation into Reference Architectures for Mobile Robotic Systems*, in: Conference on Software Engineering Advances (ICSEA), pp. 465–471, 2012.
[22] A. Ahmad, M. A. Babar, *Software architectures for robotic systems: A systematic mapping study*, Journal of Systems and Software 122, 16–39, 2016.
[23] ISO42020, *Software, systems and enterprise: Architecture processes*, ISO/IEC/IEEE 42020, 2019.
[24] L. Bass, P. Clements, R. Kazman, *Software architecture in practice*, Addison-Wesley Professional, 2003.
[25] P. Hillmann, D. Schnell, H. Hagel, A. Karcher, *Enterprise Model Library for Business-IT-Alignement*, International Conference on Software Engineering and Applications (SEAPP), AIRCC Digital Library, 2022
[26] The Open Group, *ArchiMate 3.0 Specification*, Van Haren Publishing, 2016.